\documentclass[11pt]{article}
\usepackage{geometry} 
\usepackage{graphicx}

\usepackage{array} 

\usepackage{subcaption} 

\usepackage{fancyhdr} 
\usepackage{abstract}

\usepackage{authblk}

\usepackage{eso-pic}

\newcommand{\reportnum}[2]{
  \AddToShipoutPictureBG*{%
    \AtPageUpperLeft{%
      \hspace{0.65\paperwidth}%
      \raisebox{#1\baselineskip}{%
        \makebox[0pt][l]{\textnormal{#2}}
  }}}%
}

\title{\vspace{-1cm}Towards An Updated Simulation of the Booster Neutrino Beam; Contribution to the 25th International Workshop on Neutrinos from Accelerators}
\author[1]{Josephine L Paton\thanks{jpaton@fnal.gov}} 
\date{\vspace{-0.3cm}For the SBN Program} 

\affil[1]{\small{Fermi National Accelerator Laboratory, Batavia, Illinois 60510, USA}}

\begin{document}
\maketitle
\reportnum{-4}{FERMILAB-CONF-24-0887-PPD}
\reportnum{-5}{NuFact2024-19}
\vspace{-0.6cm}

\begin{abstract}
Precise, accurate neutrino flux predictions for neutrino beam experiments are crucial for physics results. Flux predictions for the Booster Neutrino Beam at Fermilab were first published in 2009 by the MiniBooNE collaboration \cite{PhysRevD.79.072002}. It is no longer possible to run the simulation used to create these predictions due to outdated software versions. In this exciting period for the Short-Baseline Neutrino Program at Fermilab, with both the far detector (ICARUS) and the near detector (SBND) operating, an updated flux model for the BNB is necessary. For the purpose of using a dynamic, up-to-date beam model, the SBN program has created a new simulation named G4BNB. This framework contains new features, such as a full neutrino ancestry to handle hadron production systematics with more precision, and the inclusion of neutral mesons from the proton-beryllium scatter to allow the study of exotic BSM scenarios.

\end{abstract}

\section{The Booster Neutrino Beam}

The Booster Neutrino Beam (BNB) is supplied with 8 GeV protons from the Booster accelerator, and supplies neutrinos to many experiments at FNAL, including the Short-Baseline Neutrino (SBN) program. The neutrino flux predictions for the BNB were originally calculated for MiniBooNE \cite{PhysRevD.79.072002}, and subsequent experiments have rescaled these original predictions. For the SBN program, a new simulation of the BNB, known as G4BNB, has been created.

\begin{figure}[h]
\begin{center}
\begin{subfigure}{0.9\textwidth}
\includegraphics[width=\linewidth]{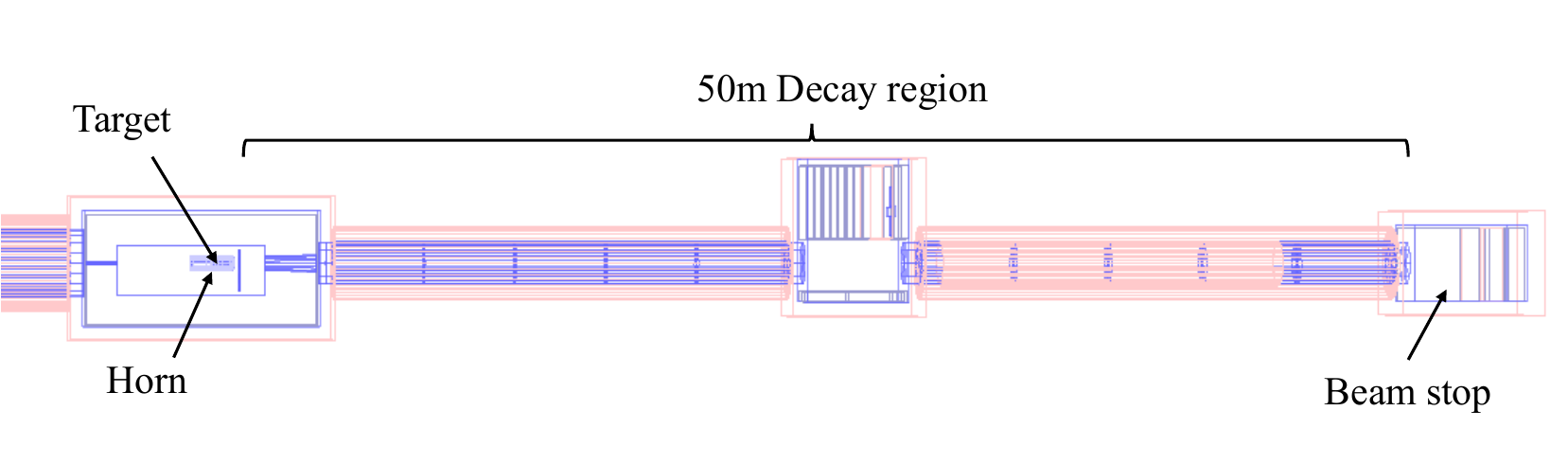}
\caption{}
\label{geo}
\end{subfigure}
\begin{subfigure}{0.55\textwidth}
\includegraphics[width=\linewidth]{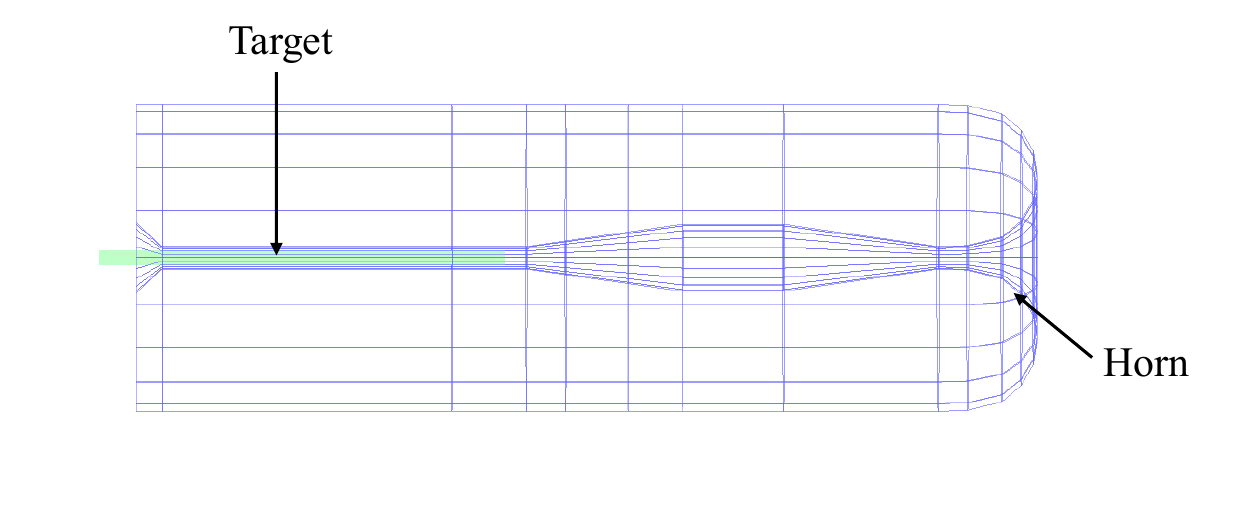}
\caption{}
\label{horn}
\end{subfigure}
\caption{A drawing of the geometry used in G4BNB. The four main components are labelled; the target, the focussing horn, the decay region and the beam stop. A view of the horn and target is shown in Subfig. \ref{horn}.\vspace{-0.5cm}} 
\label{geo_whole}
\end{center}
\end{figure}

The geometry of the G4BNB simulation was imported from the original MiniBooNE simulation, and includes a beryllium target, focusing horn, decay region and beam stop. In order to ensure the accuracy of this modeling, a survey of the BNB site has been completed in November 2024. The geometry in G4BNB v1.0 can be seen in Fig. \ref{geo}, with a closer look at the horn and target in Subfig. \ref{horn}. A detailed description of all components of the BNB can be found in Section II of \cite{PhysRevD.79.072002}.

The target is made of beryllium, comprised of several slugs held in place by fins inside an outer sleeve, with a total length of 711~mm at room temperature. The aluminum focussing horn is modeled with a variable surface current, allowing for investigation of Reverse Horn Current predictions as well as systematic uncertainty calculations. A closer look at the horn and target can be seen in Subfig. \ref{horn}. The horn is operated at a current of 174~kA with a maximum field strength of 1.5~T. The decay region is 50~m long, and ends with a beam dump made of layers of steel, concrete and air. An additional beam dump is positioned above the decay tube region at the 25~m mark, but this is not deployed into the particles' path during operation for SBN. The beam dump is comprised of layers of steel, concrete and air, with a total length of 4.48~m and a height of 3.15~m.

\section{Physics Simulation}

\begin{figure}[htbp]
\begin{center}
\includegraphics[width=0.7\textwidth]{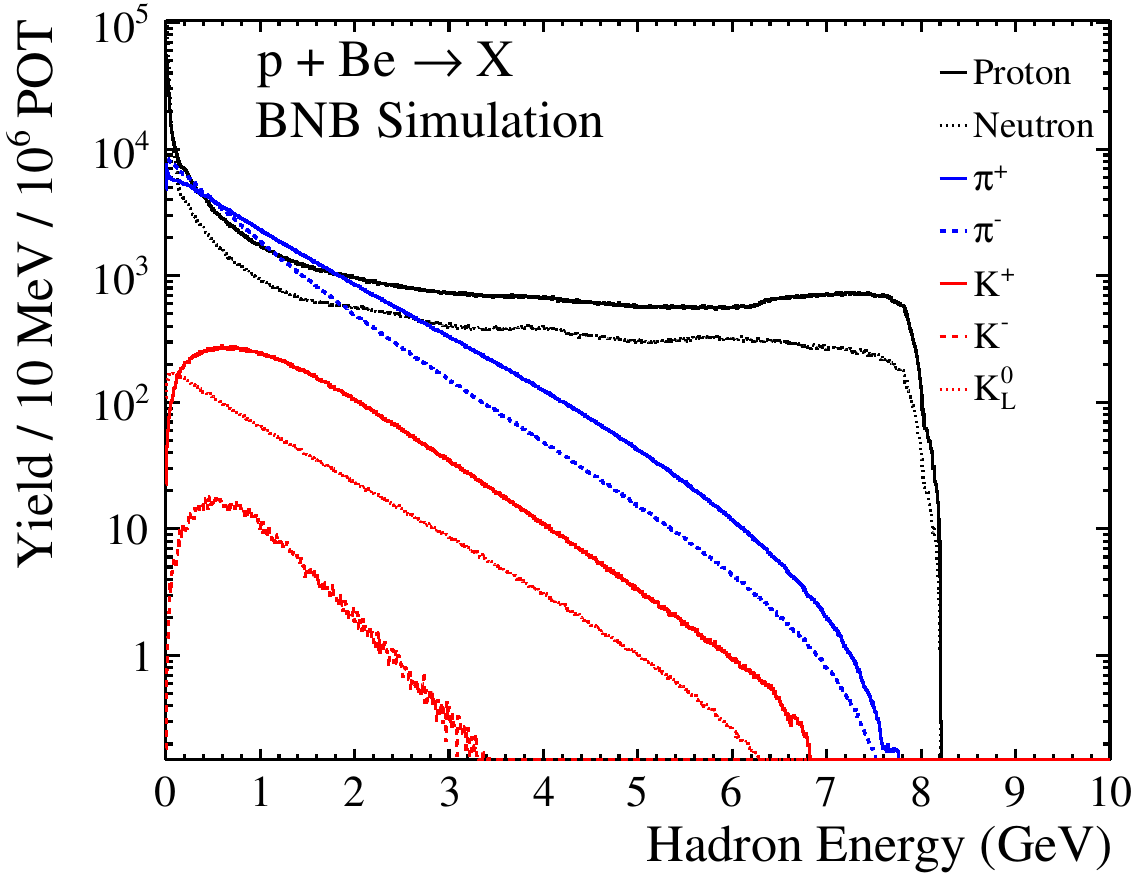}
\caption{Initial energy spectra of the products of the p-Be scatter in the G4BNB simulation. The production model used to create these particles is the same as that used by the original MiniBooNE simulation \cite{PhysRevD.79.072002}. Only particles that can originate a chain that can produce a neutrino are simulated.  \vspace{-0.5cm}}
\label{parents}
\end{center}
\end{figure}

The products of the initial proton-beryllium (p-Be) scatter in G4BNB are currently created using the same custom model as the MiniBooNE simulation. An in depth description of the custom tables created by the MiniBooNE collaboration can be found in Section V of \cite{PhysRevD.79.072002}. The energy spectra of these initial products can be seen in Fig. \ref{parents}. 

These particles are propagated using Geant4 \cite{AGOSTINELLI2003250,1610988,ALLISON2016186}. Custom physics models are used to constrain interactions of protons, neutrons and pions on beryllium and aluminum within the simulation, in line with the MiniBooNE models. The constraints are applied above 2~GeV for protons and neutrons, and above 0.5~GeV for pions. Below the constrained energy regions, the particle interactions are handled by the default Geant4 physics lists. The previous MiniBooNE BNB simulation used Geant4.8.1, which had the default QGSP BERT \cite{APOSTOLAKIS2009859}. This version of Geant4 is deprecated, meaning the simulation can no longer be sustainably used. G4BNB uses a more recent version of Geant4 so that features can be added easily and the simulation can be adjusted to investigate effects such as geometry and horn current systematic uncertainties. Geant versions since 4.10 use the default FTFP BERT \cite{ALLISON2016186}. G4BNB has currently been running with Geant4.10.4, with a view to upgrading to 4.10.6 before publication due to improved agreement with data in unconstrained regions. Due to the changes in underlying physics model, it is possible there may be differences in the flux prediction in low energy regions. With a maintained simulation that can be run on current machines, it will be possible to continue adding new models and tuning. For example, additional data sets could be used to further develop the production model, including the results of NA61/SHINE \cite{Abgrall2016} and EMPHATIC \cite{EMPHATIC:2019xmc}.\newpage

Ongoing work on G4BNB includes the addition of neutral mesons such a $\pi^0$ and $\eta$ to the production model of the initial p-Be interaction. While these particles do not initiate chains that lead to neutrinos, their decays and interactions can be used to study physics beyond the standard model. This is especially impactful at SBND, which sits only 110~m from the BNB, meaning it has the chance to observe short lived particles as well as seeing very high fluxes.

\section{Dk2Nu File Format}
 
An additional update to G4BNB is the filetype used for simulation outputs: Dk2Nu \cite{dk2nu2012}. The Dk2Nu file format saves information about the kinematics, physics process and material of every interaction in the ancestor chain leading to a neutrino. This information is invaluable in calculating accurate systematic uncertainties that result from particle interaction cross-sections - one of the largest sources of uncertainty in any neutrino flux prediction. Dk2Nu is established within the community and has been adopted by many Fermilab experiments, including DUNE, NO$\nu$A and MINER$\nu$A. Additionally, the Dk2Nu package supplies an interface to GENIE, allowing it to be used in conjunction with detector simulations. Due to adopting the Dk2Nu format, it is possible to rework the methods used for systematic uncertainty calculations within the SBN program.
\newpage

\section{Conclusions}
Flux predictions hold a crucial place in neutrino beam experiments, requiring accuracy and precision in order to reduce systematic uncertainties on physics measurements. A new prediction for the Booster Neutrino Beam flux is being prepared, using the new simulation known as G4BNB. This framework utilizes a more recent version of Geant4, can be adapted to investigate systematic uncertainties, and outputs to the established Dk2Nu file format. Further upgrades and improvements to G4BNB are planned before an eventual flux data release. 

\vspace{0.5cm}
This document was prepared by SBN using the resources of the Fermi National Accelerator Laboratory (Fermilab), a U.S. Department of Energy, Office of Science, Office of High Energy Physics HEP User Facility. Fermilab is managed by Fermi Research Alliance, LLC (FRA), acting under Contract No. DE-AC02-07CH11359.

\bibliographystyle{atlasnote}
\bibliography{proceedingsBib}

\end{document}